\documentclass[preprint,12pt,authoryear]{elsarticle}
\usepackage{amsmath}
\usepackage{wrapfig,lipsum,booktabs,longtable}
\usepackage{pdflscape}

\newcommand{\degree}{$^\circ$}
\newcommand{\angstrom}{\mbox{\normalfont\AA}}

\usepackage{amssymb}
\journal{Icarus}

\begin{document}

\begin{frontmatter}



\title{Modeling the complete set of Cassini's UVIS occultation observations of Enceladus' plume}


\author[lasp]{Ganna~Portyankina}
\author[lasp]{Larry~W.~Esposito}
\author[lasp]{Klaus-Michael~Aye}
\author[psi]{Candice~J.~Hansen}
\author[NM]{Ashar~Ali}

\address[lasp]{Laboratory for Atmospheric and Space Physics, University of Colorado in Boulder, 3665 Discovery Drive, Boulder, CO, 80303}
\address[psi]{Planetary Science Institute, Tucson, AZ}
\address[NM]{US Air Force Research Laboratory, Albuquerque, NM}

\begin{abstract}
The Cassini Ultraviolet Imaging Spectrograph (UVIS) observed a plume of water vapor spewing out from the south polar regions of Enceladus in occultation  geometry 7 times during the Cassini mission. 
Five of them yielded data resolved spatially that allowed fits to a set of individually modeled jets.
We created a direct simulation Monte Carlo (DSMC) model to simulate individual water vapor jets with the aim of fitting them to water vapor abundance along the UVIS line of sight during occultation observations.
Accurate location and attitude of spacecraft together with positions of Enceladus and Saturn at each observation determine the relationship between the three-dimensional water vapor number density in the plume and the two-dimensional profiles of water vapor abundances along the line of sight recorded by UVIS.
By individually fitting observed and modeled jets, every occultation observation of UVIS presented a unique perspective to the physical properties and distribution of the jets.
The minimum velocity of water vapor in the jets is determined from the narrowest observed individual jet profile: it ranges from 800~m/s to 1.8~km/s for the UVIS occultation observations.
41 individual jets were required to fit the highest resolution UVIS dataset taken during the Solar occultation however, an alternative larger set of linearly-dependent jets can not be excluded without invoking additional preferably unrelated data from other instruments. 
A smaller number of jets is required to fit the stellar occultation data because of their spatial resolution and geometry.
We identify a set of 37 jets that were repeatedly present in best fits to several UVIS occultation observations.
These jets were probably active through the whole Cassini mission.

\end{abstract}


\end{frontmatter}


\section{Introduction}

Early in the mission Cassini discovered a plume of water vapor and ice particles emerging from the southern pole of Enceladus \citep{Dougherty_2006}.
This was a surprising discovery, because Enceladus was at the time thought to be too small to support current geological activity.
The discovery raised broad scientific interest because it unveiled formerly unknown physical processes that operate inside the small Saturnian moon \citep{Hsu_2015, Thomas_2016, Matson_2018}, and also uncovered a possibility of detecting a habitat for extraterrestrial life \citep{Waite_2006, Postberg_2018, Porco_2017, Sekine_2015}.
Almost every instrument on board Cassini has since observed the plume gaining new knowledge on the plume's composition and activity, as well as about the under-ice reservoir that is believed to be a source for the plume \citep{Porco_2014, Hansen_2008, Hansen_2011, Postberg_2018,  Spencer_2011, Teolis_2016, Hedman_2009, Dhingra_2017}.

The south polar regions of Enceladus are geologically young and scarred by multiple grooves, faults, and fractures \citep{Barr_2008, DiSisto_2016}.
Most notable of them are large fissures informally called ''tiger stripes'' that penetrate through the icy shell of the satellite to a liquid water reservoir that feeds the plume or probably even to its liquid ocean that lies beneath \citep{Matson_2012, Kite_2016}.
The Cassini Imaging Science Subsystem (ISS) has detected nearly 100 separate collimated  particle jets set in the background of the extended plume \citep{Porco_2014}.
It is widely accepted that the sources of the jets are located within the tiger stripes \citep{Spitale_2007, Porco_2014, Spitale_2015}.
Measurements of the thermal emission support this by showing that the surface temperatures within the stripes are higher than the surrounding areas and reach at least 180~K over regions tens of meters wide along the length of the tiger stripes \citep{Spencer_2009, Howett_2011, Spencer_2011, Goguen_2013}.

The Enceladus plume provides an indirect way to study Enceladus' inner structure and composition of the subsurface reservoir that feeds the plume. 
The plume consists of the two main components: solids (micron-sized water ice particles) and water vapor.
Cassini’s remote sensing instruments that operate in visible and near-infrared wavelengths (ISS, VIMS \cite{Porco_2004, Brown_2004}) are predominantly sensitive to the solid component of the plume, while the Ultraviolet Imaging Spectrograph (UVIS, \cite{Esposito_2004}) is sensitive to its water vapor component.
For example, the 100 jets detected by ISS \cite{Porco_2014} are composed of icy grains which are lofted by the water vapor jets (the main subject of this paper).
Salt particles contribute to the solid component of the jets by $ \approx{}1 \%$ \citep{Postberg_2011}. 
In addition contributions of approximately $5\%$~CO$_{2}$, $1\%$~CH$_{4}$, $1\%$~NH$_{3}$, and small quantities of heavier hydrocarbons and heavy organic molecules were detected \citep{Waite_2009, Postberg_2009, Postberg_2018}. 

The flux of solids is observed to vary considerably depending on the orbital location of Enceladus \citep{Hedman_2013, Nimmo_2014, Ingersoll_2017}.
The total flux of water vapor in the plume is estimated to be close to 200~kg/s \citep{Hansen_2006, Hansen_2008, Hansen_2011} to 300~kg/sec \citep{Hansen_2019}. 
It is observed to be stable over the course of Cassini mission with no more than 20 percent variability, however, UVIS sees variations in the fluxes from collimated gas jets on a diurnal scale  \citep{Hansen_2017}.
\cite{Kite_2016} hypothesized that while the tidal forces modulate openings of the fissures, different fissures open differently along the Enceladus orbit. 
UVIS data show vapor flux production from different jets modulates along the orbit on a diurnal cycle, while the total flux does not vary significantly. 
Consequently, different jets thus have different lifting forces for solid particles. 
While we have a reasonable estimate of the total gas production of the jets, the total mass of icy grains is somewhat uncertain: the estimates of ice/gas ratio range from  0.2 up to 1, with probably the most accepted range being 0.35 to 0.70 \citep{Porco_2006, Kieffer_2009, Hedman_2009, Ingersoll_2011}.
The ratio of ice particle to vapor mass is indicative of transport and formation of ice particles \citep{Ingersoll_2011} and provides information on irregularity of sub-surface conditions underneath southern polar terrain \citep{Hedman_2018}. 
The flux of solids is observed to vary considerably depending on the orbital location of Enceladus \citep{Hedman_2013, Ingersoll_2011, Ingersoll_2017} which is compatible with the hypothesis of \cite{Hansen_2017}.

In this work we look at all the occultation observations conducted by UVIS.
We use a direct simulation Monte Carlo (DSMC) model to fit single jets to water vapor abundance along the UVIS line of sight (LoS) at each point of the observations. 
We compared the outcomes with the aim to estimate which jets remain active from observation to observation and which, in contrast disappear.

This paper is organized as following: section \ref{sec:observations} introduces the data acquisition strategy of UVIS while using solar occultation data as an example, section \ref{sec:model} postulates the assumptions and set-up of the DSMC model that we use to fit UVIS observations and touches on the fitting procedure. We use Solar occultation here as an example again, finalizing the complete treatment procedure for a given occultaiton observation.
Section \ref{sec:observations} summarizes the whole set of the occultation observations conducted by UVIS and the best fits obtained with the use of the model.
We finish with conclusions and discussion in section \ref{sec:discussion}.

\section{Observational data: overview of UVIS occultation observations during Cassini mission}\label{sec:observations}

UVIS observes Enceladus' jets in stellar and solar occultations when a star passes behind Enceladus' plume as seen from the spacecraft.
While Enceladus and its atmosphere occult a star, UVIS takes spectroscopic measurements in its Far UltraViolet (FUV), and High Speed Photometer (HSP) channels (the 2010 Solar occultation was an exception to this, when UVIS used the Extreme UltraViolet (EUV) channel instead).
In the UVIS occultation mode solar channel the spatial dimension of the solar/stellar image is projected onto two windows on the detector, and focused onto 1024 pixels in the spectral dimension, each having a width of 0.606 \angstrom{}̊ \citep{Esposito_2004}.

Table \ref{tbl:occs_data} lists all the occultation observations of Enceladus plume conducted by UVIS through the Cassini mission.
Columns 3 and 4 list properties important to application of the DSMC model: the water vapor column number densities and total water vapor flux derived by \cite{Hansen_2017} from UVIS data.
UVIS had 7 occultation observations, with one of them, in 2011, being a double occultation -- 2 stars were crossing the jets simultaneously at different heights relative to Enceladus' limb.

\begin{landscape}
\begin{table}[h]
\begin{tabular}{ | l | l | l | l | l | l |}
\toprule
Date 				& Occulting star 					& I0  								& Column number density 			& H$_2$O Flux   	& Mean\\
						& 											&	interval							& $n \times 10^{16} cm^{-2}$ 		& [kg/s]					& Anomaly\\
\midrule
17 Feb 2005 	& $\lambda$ Scorpii 			& 	--									& no detection 								& no detection 	& --		\\
14 Jul 2005 		& $\gamma$ Orionis 			& 	NA								& 1.5 $\pm$ 0.15			& 240 			& 117	\\
24 Oct 2007 	& $\zeta$ Orionis 					& 16:59:49 -- 17:00:49 	& 1.4 $\pm$	0.14			& 338			& 236	\\
18 May 2010 	& The Sun 							& 05:54:36 -- 05:55:01     & 0.9 $\pm$	0.23			& 337 			& 98		\\
19 Oct 2011 	& $\epsilon$ Orionis 			& 09:21:00 -- 09:21:50     &1.35 $\pm$ 0.15			& 313			& 237	\\
						& $\zeta$ Orionis 					& 09:21:00 -- 09:21:50		&1.2 $\pm$ 0.2			& 313 			& 237	\\
11 Mar 2016		& $\epsilon$ Orionis 			& 11:51:25 -- 11:52:00     & 1.5 $\pm$ 0.15 			& 250 			& 208	\\
27 Mar 2017 	& $\epsilon$ Canis Majoris 	& 14:38:30 -- 14:39:10      &1.0 $\pm$(?) 				& 275 			& 131	\\
\bottomrule

\end{tabular}
\caption{\textbf{\label{tbl:occs_data}} Complete set of UVIS occultation observations conducted by UVIS. Number densities and fluxes are from \cite{Hansen_2019}. }
\end{table}
\end{landscape}

In the following subsections, we will use UVIS data from the Solar occultation on 18 May 2010 as an example for data treatment and model application.
We will discuss the remaining occultations and model fits in the subsections \ref{sec:GaOri2005} - \ref{sec:ZeOri2007}.

\subsection{Example of data processing: Solar occultation, 18 May 2010}\label{sec:solar_occ}

\begin{figure}[h!]
\begin{center}
\includegraphics[width=0.98\columnwidth]{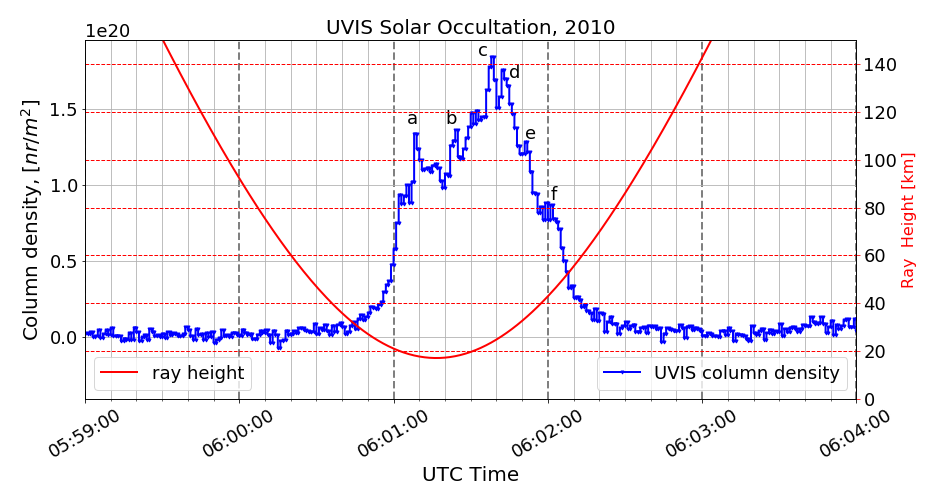}
\caption{\textbf{\label{fig:solar_occ_data}}
The column water vapor density (blue curve) as derived by UVIS Solar occultation observation on May 18, 2010 and the ray height (red curve) of UVIS LoS during the occultation. Letters mark the most prominent absorption features that roughly correspond to locations above the tiger stripes.}
\end{center}
\end{figure}

On May 18 2010, Enceladus plume, the Sun and the Cassini spacecraft were in positions to allow UVIS acquire solar occultation data \citep{Hansen_2011}.
We used the UVIS EUV channel with 1~sec integration time and detected attenuation of the solar light by water vapor in 95 successive spectra.
We can calculate the attenuation of the solar flux by the plume from the UVIS spectrum by summing over the wavelength range 850--1000~ \angstrom{}̊ and normalizing the occulted signal by the non-occulted signal.
Doing so for every spectra successively acquired during the occultation we can plot I/I0 -- or effectively water vapor optical depth -- as a function of time.
To increase the signal-to-noise ratio for the non-occulted signal I0, we routinely sum over several minutes of UVIS signal before or after the star signal is reduced by the jets.
The result of this calculation is shown as a green curve in Figure \ref{fig:solar_occ_data}.
The red curve is the ray height of the UVIS' LoS during the occultation with vertical axis on the right hand side of the plot.

The solar occultation has the highest temporal resolution of all UVIS occultation observations and thus allows detection of the largest number of separate jets.
The geometry of this occultation was beneficial for the purpose of identifying separate jets.
Most importantly, the minimum ray height of the LoS (i.e. the closest distance of the LoS to Enceladus’ surface) was below 60 km altitude above the surface of Enceladus for almost 2 minutes.
UVIS sensed the deepest parts of the plume providing the highest signal to noise ratio for water vapor abundances allowing  the detection of its small variations.  
In addition, the ground track of the observation was almost but not quite perpendicular to tiger stripes (Fig. \ref{fig:solar_fit_map}), which means that the LoS of UVIS crossed tiger stripes at a slight angle.
This allowed for minimum overlap of the jets from different tiger stripes at each observational point and thus using the model we are able to distinguish the maximum number of individual jets.

The optical depth can then be converted to the water vapor column density along the LoS of UVIS.
The blue curve in Figure \ref{fig:solar_occ_data} indicates a column density along the LoS of UVIS plotted vs. relative ephemeris time (et) during the observation.
The narrow collimated jets can be identified as enhanced absorption features.
While we plot here the result from one spatial window of UVIS, the enhanced absorption features do appear in both UVIS spatial windows, ruling out shot noise \citep{Hansen_2011}.

\begin{figure}[h!]
\begin{center}
\includegraphics[width=0.98\columnwidth]{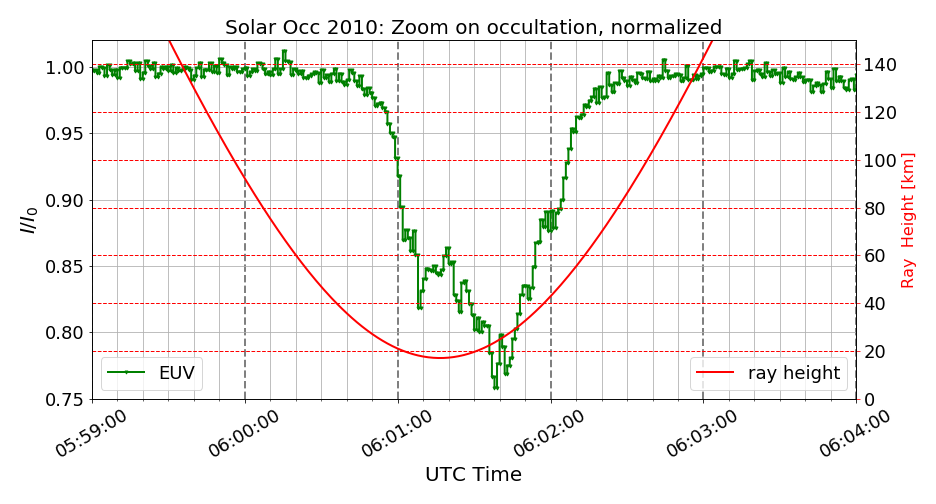}
\caption{\textbf{\label{fig:Solar2010_data}}
UVIS solar occultation on 18 May 2010: the green curve is the occulted UVIS signal in the EUV channel summed over 850--1000 \angstrom{} normalized by the unocculted signal. The red curve shows the ray height of the UVIS LoS during the occultation. } 
\end{center}
\end{figure}

The relative positions of the observed star (or the Sun), Enceladus and the spacecraft and the attitude of Cassini spacecraft  constantly change during the observation.
This means that for each point in Figure \ref{fig:solar_occ_data} the optical path of the LoS of UVIS through the plume is slightly different.
The red curve in Figure \ref{fig:solar_occ_data} shows the ray height for each corresponding observational point.
Generally, the occulted signal produced by the water vapor in the jets is detectable starting from 06:00:45.
One can see that the ray height has its minimum at 06:01:19, while the maximum absorption is at approximately 06:01:40.
This means that the part of the jets that UVIS senses at deeper altitudes at 06:01:19 is less intense than those it senses at later time and altitude.
This observation serves as a motivation to do more in-depth investigation into geometrical relationships between single jets and the LoS of the UVIS instrument.

\section{Model}\label{sec:model}

\subsection{DSMC model for jets}

We have constructed a 3D direct simulation Monte Carlo (DSMC) model for Enceladus' jets and we apply it to the analysis of UVIS occultation observations.
The current model is a further development of the model described in \cite{Tian_2007}.

\begin{figure}[h!]
\begin{center}
\includegraphics[width=0.5\columnwidth]{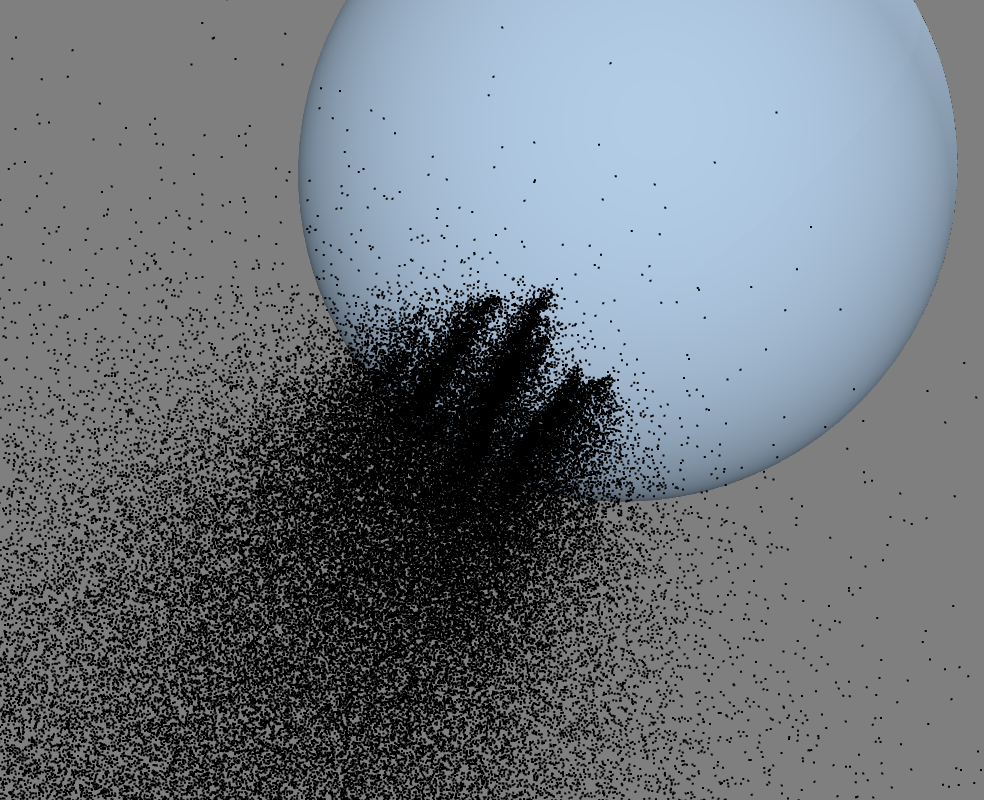}
\caption{\textbf{\label{fig:model_run}}
An example of a model run that included all 100 jets from \cite{Porco_2014}. Each black dot here represents 1e3 test particles for convinience of plotting. Each of the test particles is scaled to the total water vapor flux derived by \cite{Hansen_2017} at the step of model fitting. }
\end{center}
\end{figure}

The Monte Carlo model tracks test particles from their source at the surface into space.
In this paper we consider localized sources resembling isolated jets.
In the first case the initial positions of all test particles for a single jet are fixed to one of 100 jet sources identified by \cite{Porco_2014} (shown in Figure \ref{fig:model_run}).

The initial three-dimensional velocity of each test particle contains two components: a velocity V$_z$ that is perpendicular to the surface, and a thermal velocity V$_{th}$ which is isotropic in the upward hemisphere.
The direction and speed of the thermal velocity of each test particle is chosen randomly but the ensemble moves isotropically at a speed which satisfies a Boltzmann distribution for a temperature T$_{th}$.
The selection of a particular value for T$_{th}$ is directed by the measurements of surface temperatures in the direct vicinity and inside the tiger striped by the CIRS instrument \citep{Howett_2011, Spencer_2009, Spencer_2011}.
The emission from the brightest regions of tiger stripes observed by CIRS is close to a blackbody with best-fit temperatures in the range 167 -- 185 K.
The values for corresponding thermal velocities thus are V$_{th}$ 391 -- 412 m/s.
If not mentioned otherwise, we use a value in the upper side of this range T = 180 K and the corresponding thermal temperature of 407 m/s.

V$_{th}$ is the parameter determining the jet widening rate per altitude and as such directly determined by the gas temperature.
A range for reasonable V$_z$ is then determined by requiring that modeled jet widths match the observed ones.
We determine the V$_z$ for each occultation observation separately based on the slimmest feature (i.e jet) observed during that occultation. 
Thus value of V$_z$ simultaneously reflect the collimation degree of the jets and temporal resolution of UVIS data.  
The degree of collimation of a jet is physically determined by the processes inside the conduit below the surface.
However, that information is smoothed by the DSMC model in its current shape because the only relevant velocity values are those that test particles possess right above the surface when they exit the conduit.
In this work we only consider jets with V$_{z}$ perpendicular to the local surface (i.e. we do not make use of jet inclination information provided by \cite{Porco_2014}).
However, total velocity vector of a single test particles can be inclined due to randomly inclined thermal component. 

The model uses no grid structure to keep track of the test particles' position and velocity vectors.
However, similar to gridded models it uses a fixed time step that is chosen based on the test particles maximum vertical velocity $V_z+V_{th}$ and the desired model resolution $L_{res}$, i.e. it is calculated as:

\begin{equation}
dt = \frac{L_{res}}{V_{z}+V_{th}}
\end{equation}

To compare results of the model simulation to UVIS data the model's spatial resolution must be higher than the smallest changes of the LoS ray height between 2 consequent steps in time during the observation.
Effectively, this means that $L_{res}$ depends on temporal resolution of UVIS data and distance of the spacecraft to Enceladus jets during the observation.
For example, for the solar occultation from May 2010 this is equal to 1~km while for the stellar occultation of $\epsilon$  Orionis from  2011 it is 20~km.

The positions and velocities of the test particles are influenced by gravity of Enceladus and Saturn as well as Coriolis, centrifugal, and tidal forces, and by test particle collisions.
Collisions are treated statistically: first, at each time step for every test particle the model evaluates the mean free path from the gas density in test particle's immediate vicinity.
The surrounding volume that is evaluated depends on the model resolution and thus indirectly by the temporal resolution of the observation under consideration.
By comparing the mean free path to $L_{res}$ the model determines if the test particle is in a collisional or collisionless regime.
In a collisionless regime the test particle velocity vector is not altered, otherwise the Bernoulli trial method is used to determine if the test particle experienced a collision or not: a random number is drawn and compared to a probability that the test particle would experience a collision at the calculated gas density \cite{Papoulis_1984}.
If the test particle experiences a collision, its velocity is altered: it is given a new velocity vector drawn from the Boltzmann distribution with the mean velocity of the test particles around it.

All test particles reaching 3 satellite radii distance or hitting Enceladus's surface are replaced with test particles from the source to maintain the total number of test particles in the model domain.

During each simulation run, the model is first required to reach steady-state, as defined by minimal variations of the 3-dimensional number density of test particles in the whole model domain.
A precise criterion for steady-state depends on the number of test particles in the particular run and $L_{res}$.

In general terms the use of a steady-state condition is justified for physical processes that are unchanging in time.
The height of the LoS of the UVIS instrument at times during the observation at which it detects the signal from the plume is of order of 50~km.
If a test particle has velocity in order of 200~m/s (lower limit) then it takes this particle 4 minutes to reach the level at which it could be detected during the observation.
Currently we have no reason to assume fluctuations of Enceladus jets eruptions at this or higher frequencies.
Another slightly different precaution for the steady-state must be considered for the total length of the observation.
At each point of time during the occultation the LoS of the instrument crosses different jets.
It is possible to think of a situation where UVIS looks through a jet and it turns off while it is still being observed but in a slightly different geometry, i.e. at a later time slot.
Utilizing steady-state solution here means that the jets are steady during the whole observation time while the plume signal is detected, i.e. 95 seconds for Solar Occultation in 2010, 10 seconds for $\zeta$ Orionis 2007, 27 seconds -- for  $\epsilon$ Orionis 2011 and so on.

After the model run has reached the steady-state, the test particle flux is calculated to scale the test particles with observed water vapor production rates \citep{Hansen_2017}.
At this point we consider all the modeled jets producing equal amounts of water vapor and we scale all the test particles equally.
Later the production rates of the jets will be adjusted using the fitting procedure \ref{sec:fitting}.
This is mathematically equivalent to applying different scaling to test particles from different jets.
\citep{Hansen_2019} updated the total water vapor flux number to 300~kg/s from 200~kg/s that was published previously \citep{Hansen_2017}.
We have used the old number in this work.
The increase in the observed water vapor flux means an increase in the scaling factor for a test particle in our model.
It does not affect the relative strengths of the jets which are the prime topic of this paper.

Each model run results in a set of coordinates and velocities of a given set of test particles in the reference frame of Enceladus.
Figure \ref{fig:model_run} shows a 3-D rendering of a resulting steady-state solution for a set of 100 singular jets where each black dot is a test particle.
These are converted to the test particle number densities and then integrated along the UVIS LoS for each time step of the UVIS occultation observation using the SPICE system \citep{Acton_1996}.

The geometry of each observation, i.e. the positions of the spacecraft, orientation of the camera, position of Enceladus, and the rotation of the jets toward or away from the spacecraft are calculated at every time step of each observation.
The 3-D vector of the LoS is derived relatively to the instantaneous position of Enceladus and then used in the model to calculate the cumulative density of test particles along the LoS.
This calculation is repeated for all time steps of the observations.
The resulting curve is a modeled cross-section of each jet (or a broad plume, or curtain plume, depending on the modeled source) as it would be seen by UVIS.

\subsection{Fitting procedure}\label{sec:fitting}

This section describes the fitting procedure we used to adjust the relative strength of jets from a set of 100 jets described in \cite{Porco_2014}.
Similar procedures can be used for curtain-type sources if it is run as a set of elongated sources along a tiger stripe and not as a single source.

\begin{figure}[h!]
\begin{center}
\includegraphics[width=0.95\columnwidth]{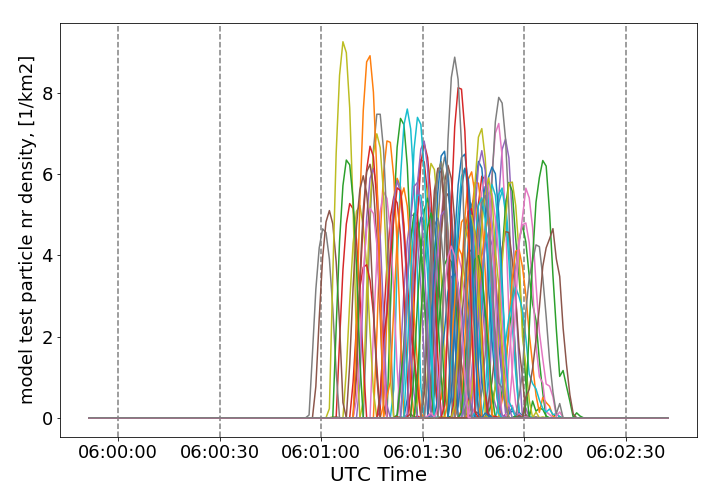}
\caption{\textbf{\label{fig:all_jets}}
Profiles of all 100 jets from \cite{Porco_2014} as they would have been seen by UVIS during the Solar occultation observation in 2010.}
\end{center}
\end{figure}

In the case of the isolated point sources, the main result of the simulation run is a test particle number density along the UVIS LoS for each time point during the occultation observation for each of the jets separately.
Each jet results in a curve with time on the x-axis and test particle number density on the y-axis.
If we plot all the 100 jets as if we would be able to see them separately, the outcome will resemble Figure \ref{fig:all_jets}.
Individual jets resemble Gaussian curves, however, they are not necessarily Gaussian curves and can be asymmetric depending on how the LoS of UVIS crosses the modeled jet.
In reality, the geometry of the occultation observation is such that the LoS at each point of time crosses multiple jets from different tiger stripes.
Hence at each point of time contributions of all jets that are crossed by the LoS must be summed to be compared to the UVIS opacity.
This leads to some features being linear combination of multiple jets.
Thus, the first step in the fitting procedure is to reduce the original  set of all modeled jets to a linearly independent set.
By the nature of this procedure multiple combinations of the jets can constitute the resulting linearly independent solution set.
The number of linearly independent jets is a function of observational geometry, model resolution, and single jet geometry that mostly results from test particles velocities.
It does not depend on the actual observed optical opacity of the individual jets, which is fitted later to the DSMC model.

\begin{figure}[h!]
\begin{center}
\includegraphics[width=0.95\columnwidth]{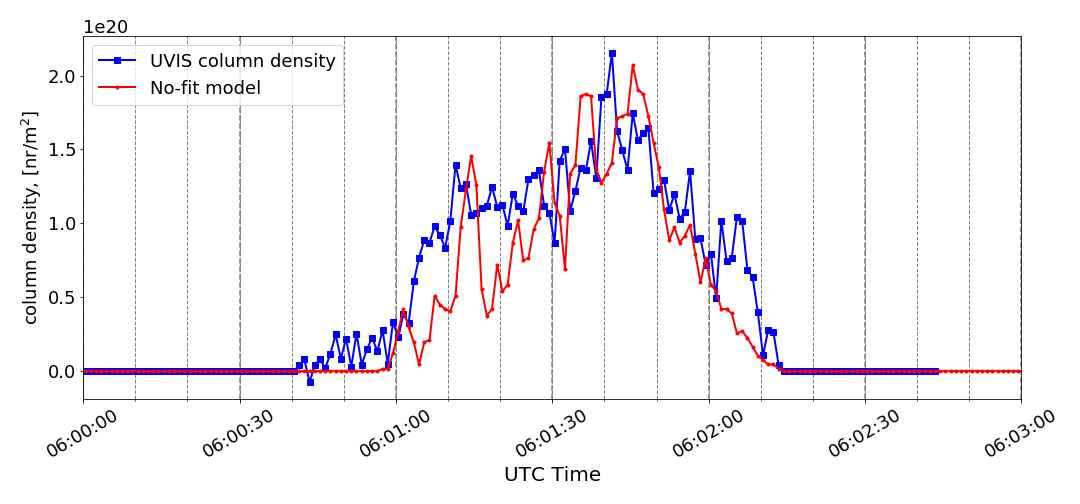}
\caption{\textbf{\label{fig:no_fit_curve}}
No-fit model for the Solar occultation observation in 2010 with $V_{th}$=1.8~km/s (red curve). All 100 jets have coefficients k$_i$ = 1. Water vapor derived from UVIS observations shown as a blue curve. The general shape of the modeled curve is consistent with the observation while local enchantments along the LoS must be modified to better fit the data.}
\end{center}
\end{figure}

The relative strength of the jets can be modified by multiplying each of them by a coefficient $k_i$.
Figure \ref{fig:no_fit_curve} shows the summed curve for model of the solar occultation if all the $k_i$=1.
It is easy to see that the relative strength of the jets must be adjusted to fit the observed UVIS occultation observation plots.

We consider the case with 100 modeled jets and $m$ time points $t_i$ in the UVIS observation.
In other words we have a $100 \times m$ matrix of values for modeled number density for each separate jet at each moment of time.
Each model jet is a column in this matrix $J$:

\begin{equation}
\begin{vmatrix}
J_{t_1, 1} 	& J_{t_1, 2} 	& .. 	& J_{t_1, 100} 	\\
J_{t_2, 1} 	& J_{t_2, 2} 	& .. 	& 	 			\\
J_{t_3, 1} 	& .. 			& ..	 	&  				\\
.. 			& 				& 		&  				\\
.. 			& 				& 		& 				\\
J_{t_m, 1} 	& .. 			& .. 	& J_{t_m, 100}
\end{vmatrix}
\end{equation}

Then we performed Gaussian elimination on $J$ using the three elementary row operations resulting in the reduced row echelon form $(RREF)$ of $J$.
The RREF of a matrix is always unique and among other properties, it can tell us immediately if the columns of $J$ are linearly independent or not.
If not, then the $RREF$ can label which columns are linearly independent as well as how the rest can be written as a linear combination of the linearly independent set \citep{Kreyszig_2006}.

For example, looking at $RREF(J)$ from the model run done for the solar occultation geometry with 100 jets with $V_z=1.8~km/s$, we saw that only $N_{ind} = 72$ of the jets were linearly independent and the rest of the 28 could be written as a linear combination of the 72 jets.
This means that 28 of the jets contain no new information and can be discarded without any loss of information.

To fit it to observed UVIS number densities we have to solve an equation that in matrix form can be written as:

\begin{equation}
\begin{vmatrix}
J_{t_{1,1}, 1} 	& J_{t_{1,1}, 2} 	& .. 	& J_{t_{1,1}, N_{ind}} 	\\
J_{t_{2,1}, 1} 	& J_{t_{2,2}, 2} 	& .. 	& 	 			\\
J_{t_{3,1}, 1} 	& .. 			& ..	 	&  				\\
.. 			& 				& 		&  				\\
.. 			& 				& 		& 				\\
J_{t_{m,1}, 1} 	& .. 			& .. 	& J_{t_{m,1}, N_{ind}}
\end{vmatrix}
\begin{vmatrix}
k_1 		\\
k_2		\\
k_3		\\
.. 		\\
.. 		\\
k_{N_{ind}}
\end{vmatrix}=\begin{vmatrix}
Q_{t_{1,1}} 		\\
Q_{t_{2,1}}	\\
Q_{t_{3,1}}		\\
.. 		\\
.. 		\\
Q_{t_{m,1}}
\end{vmatrix}
\label{eq3}
\end{equation}

Considering the linearly independent set, we perform constrained linear least squares fitting to write the UVIS measurements as a combination of these jets with varying strengths.
The constraints were that all of the $k_i$ coefficients must be non-negative.
We can use any constrained optimizer for this problem.
Since the problem is quadratic, there is guaranteed to be a global minimum (which may not be unique because of the constraints).
Therefore the choice of the algorithm doesn't really matter.
The algorithm we used for our current problem was the "trust-region-reflective" algorithm described in \cite{Coleman_1996}.

All of the negative values in the UVIS data were reset to zero and then the UVIS data was normalized for numerical stability.
This is why for stellar UVIS occultation observations in section \ref{sec:other_data} we do not plot absolute column density but adhere to the normalized number densities. 
It can be reverted to absolute values for water vapor column density using data from Table \ref{tbl:occs_fits}.

\begin{figure}[h!]
\begin{center}
\includegraphics[width=0.98\columnwidth]{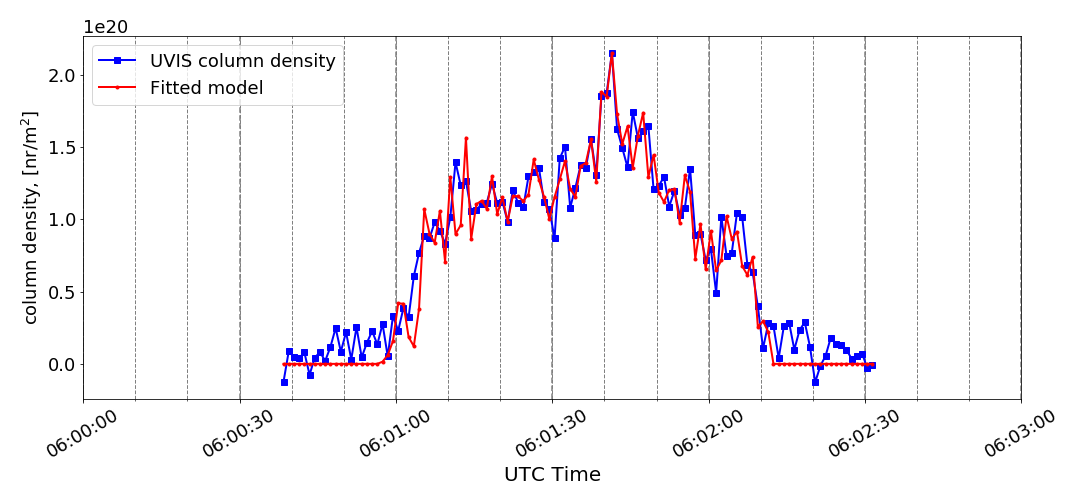}
\caption{\textbf{\label{fig:solar_fit}}
DSMC model fit to the UVIS solar occultation data.  }
\end{center}
\end{figure}

For the solar occultation example, we are considering now only the 72 jets to perform constrained linear least squares to write the UVIS measurements as a linear combination of the 72 jets.
For the sake of computational efficiency we have excluded un-occulted data from the data used in the fitting procedure and normalized the occulted signal.
After obtaining a fit, we can scale the acquired jets to the total water vapor flux calculated by \cite{Hansen_2017} and over-plot it with UVIS column density as shown in Fig. \ref{fig:solar_fit}. 
After obtaining the best fit coefficients for all 100 jets, we saw that some of them were extremely small -- between 1e-7 and 1e-12 -- and thus those jets contributed very little.
Coefficients less than 1e-7 were set to zero resulting in an additional 31 of the coefficients being zero.
In the end, the UVIS vector for the solar occultation can be written as a linear combination of only 41 model jets with vertical velocity of the water vapor of  $V_z=1.5~km/s$.
The red curve in Figure \ref{fig:solar_fit} shows the resulting fit of the 41 model jets to the observed UVIS occultation data.
One can see that the fit successfully reproduces most of the variations in UVIS column density.

\begin{figure}[h!]
\begin{center}
\includegraphics[width=0.98\columnwidth]{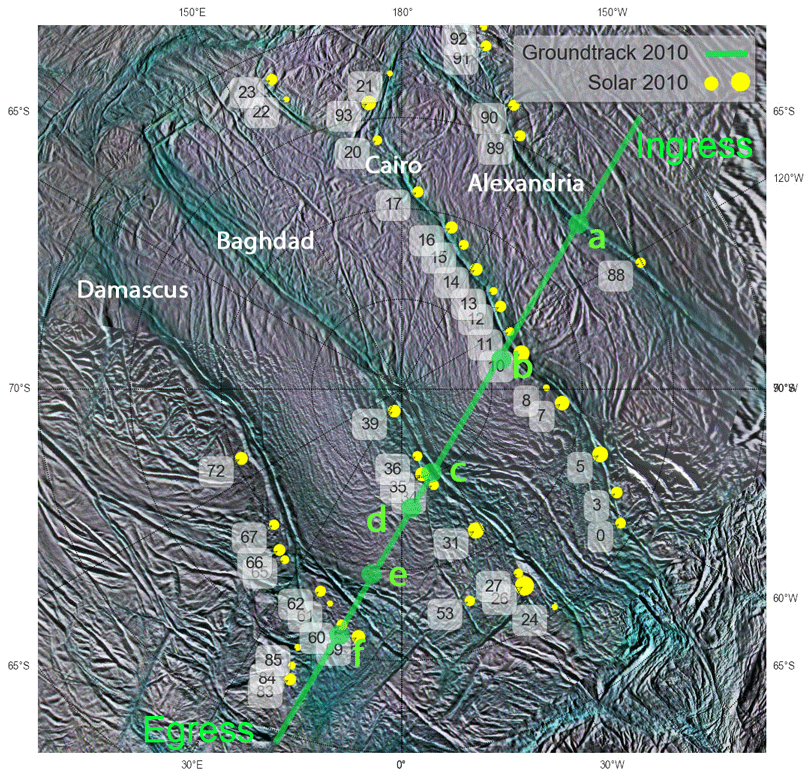}
\caption{\textbf{\label{fig:solar_fit_map}}
Locations of 41 jets that contribute to the best fit of UVIS data plotted over southern polar terrain map of Enceladus (basemap by P.~Schenk \citep{Schenk}).  Labels correspond to the numbering of the jets in \cite{Porco_2014}. The size of circle reflects relative strengths of the jets determined by the value of fit coefficients. }
\end{center}
\end{figure}

The corresponding source locations for the 41 modeled jets that contribute to the best fit are shown in the map of southern polar regions in Fig. \ref{fig:solar_fit_map}.
The shown fit theoretically is not a unique solution because of the existence of the linearly dependent sets of jets. 
However, in the case of this particular occultation this is compensated by the geometry of the observation: the LoS was hardly crossing jets from different tiger stripes at the same moment of time. 
This means that confusion between the jets starting form different tiger stripes is rare. 

\section{Observations and fitting results}\label{sec:other_data}
\subsection{Early occultations:  $\lambda$ Scorpii and $\gamma$ Orionis, 2005}\label{sec:GaOri2005}

In 2005 UVIS conducted two occultation observations.
The first occultation was of the star $\lambda$ Scorpii.
The star passed behind Enceladus disk intercepting it in equatorial region: at latitude 15\degree{} at ingress and -30\degree{} at egress.
UVIS did not detect water absorption features in this observation, neither before nor after the star was blocked by the moon's disk, thus providing limits on the latitudinal extent of the jets \citep{Hansen_2008}.

\begin{figure}[h!]
\begin{center}
\includegraphics[width=0.98\columnwidth]{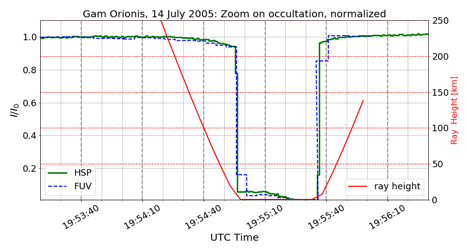}
\caption{\textbf{\label{fig:2005_data}}
UVIS occultation observation of $\gamma$ Orionis on 14 July 2005. }
\end{center}
\end{figure}

On 14 July 2005 UVIS observed the star $\gamma$ Orionis while it approached southern pole of Enceladus and intercepted the disk at latitude -76\degree{}. 
In this geometry the star cut through Enceladus' jets which is evident by the UVIS detection of attenuation of the star light by the water vapor in the plume \citep{Hansen_2008}.
The FUV channels was used with 5~sec integration time and the HSP -- with 2~msec.
Fig. \ref{fig:2005_data} shows the recorded data.
The FUV signal is summed over all wavelengths and spatial pixels  to produce signal vs time and then normalized to the un-occulted mean signal around UTC~=~19:53:03 (blue curve) when no contribution of the jets is expected to influence the signal.
The HSP signal is binned to 1~sec intervals and normalized to the un-occulted signal at the same time interval (green curve).

Attenuation of the signal is visible in the data of both channels from the time stamp of approximately 19:54:40 and until the star passed behind the limb at 19:54:57.
This provides us with 17 data points from the HSP and 3 data points from the FUV channel that show data that can be firmly attributed to the opacity of the jets.
The geometrical layout of this occultation is steeply vertical: at 19:54:40 UVIS LoS crossed the jets at altitude of 100~km and after only 17~sec it hit the hard limb.
We have found this data to not be suitable for our fitting procedure, and the system (\ref{eq3}) to be largely under-determined.

\subsection{$\zeta$ Orionis, 2007}\label{sec:ZeOri2007}

In 2007, UVIS observed an occultation of $\zeta$ Orionis.
This is the first occultation in which UVIS recorded a profile through the complete plume and revealed supersonic gas jets  \citep{Hansen_2008}.
\cite{Hansen_2008} detected 4 water vapor jets and traced them to origins of dust jets located on tiger stripes determined from ISS data \citep{Spitale_2007}.
The narrowness of the absorption features observed by UVIS argues in favor of water vapor escaping Enceladus at supersonic velocity.

\begin{figure}[h!]
\begin{center}
\includegraphics[width=0.98\columnwidth]{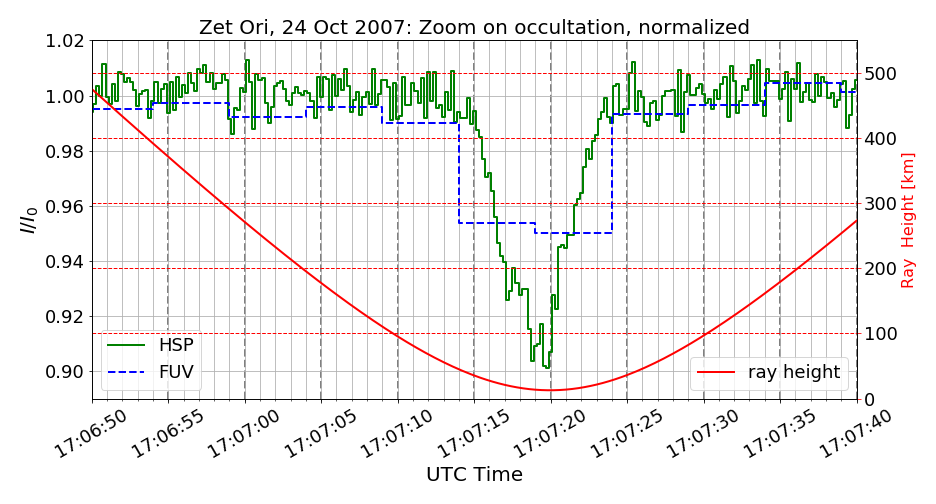}
\caption{\textbf{\label{fig:2007_data}}
HSP and FUV data from the UVIS occultation observation of $\zeta$ Orionis on 24 October 2007.  The red curve shows the ray height of UVIS LoS.}
\end{center}
\end{figure}

UVIS used the FUV channel and HSP for this observation (Fig. \ref{fig:2007_data}).
The FUV channel had 5 seconds integration time (blue curve on Fig. \ref{fig:2007_data}), which provides two points across the whole plume and renders this data useless for the purposes of modeling fits, yet it has provided information about the composition of the plume.
The HSP however was used with 2~ms integration time thus providing sufficient time resolution along the LoS to resolve separate jets with the DSMC model.
To increase the signal-to-noise ratio we have binned HSP data to 0.2~sec intervals (Fig. \ref{fig:2007_data}, green curve) and used this profile for the fitting procedure as described in Section \ref{sec:fitting}.
Attenuated signal spans from UTC~=~17:07:14 to 17:07:24, providing us with 50 significant data points to attempt the model fit.
This means that we can not attempt to fit the complete set of 100 jets to these data.
Instead, we determine the linearly-independent set of jets based on the observation's geometry and only use those jets to perform the fit.

\begin{figure}[h!]
\begin{center}
\includegraphics[width=0.45\columnwidth]{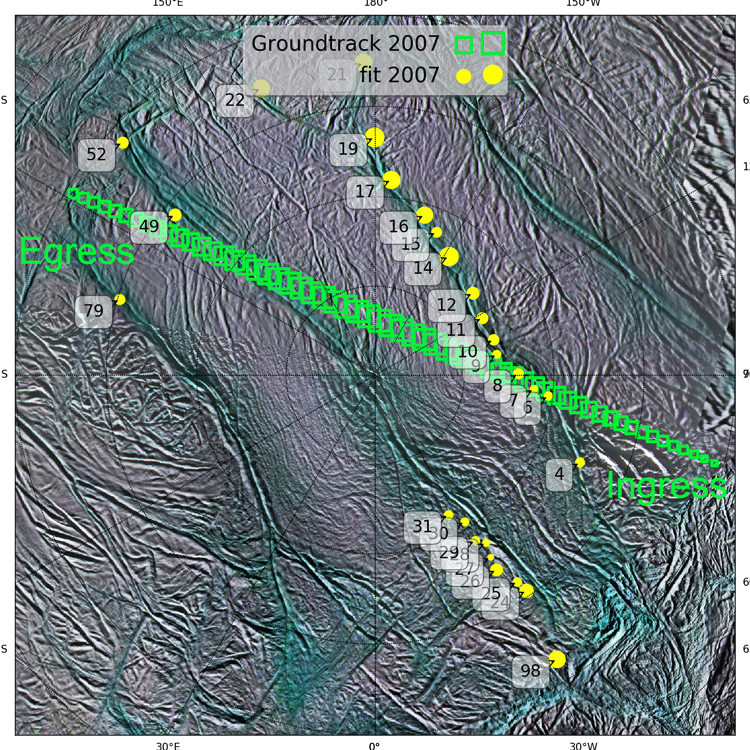}
\includegraphics[width=0.45\columnwidth]{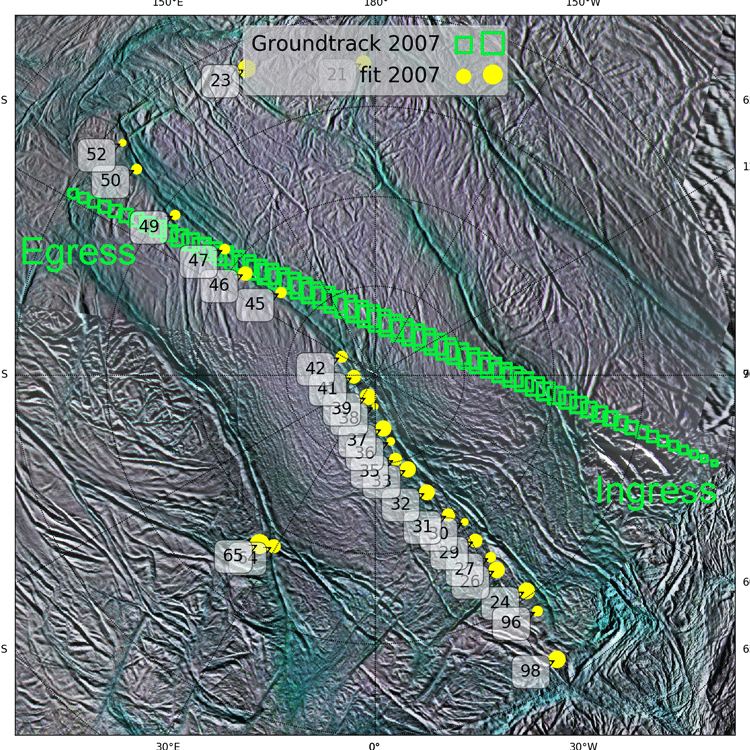}
\caption{\textbf{\label{fig:2007_maps}}
Locations of jet sources from two linearly independent sets of jets. Both sets constitute a fit to the UVIS occultation observation of $\zeta$ Orionis on 24 October 2007 (shown in the right panel of Fig. \ref{fig:2007_data}). Data from different instruments could provide useful insights for differentiation between the right and left sets of jets. }
\end{center}
\end{figure}

The geometry of this observation can be accessed via the minimum ray height and ground track of the LoS of the UVIS instrument.
The closest distance of the LoS to the surface happened at 17:07:20, with RH~=~15.6~km and coincided with the highest attenuation of the HSP signal of about 10\%.
This is different from the above-mentioned Solar occultation where the strongest absorption did not coincide with the smallest ray height. 
The ground track is the location of the interception point between a perpendicular dropped from the LoS to the surface and the surface of Enceladus.
We also plot it on polar maps (Fig. \ref{fig:2007_maps}) with rectangles and note that their relative size is proportional to the ray height at that same moment.
The geometry of this occultation is very different from the solar occultation: at each point of time the UVIS LoS crosses jets from several if not all tiger stripes.
This means the linear dependency in the set of jets is expected to be high.

We have run the DSMC model with $V_z$~=~1.2~km/s equal for all jets with all the jets perpendicular to the ground. 
The selection of vertical velocity is determined by the need to fit the narrowest absorption features in the UVIS data.
We have determined the lowest velocity by directly comparing the width of these features to the cross-sections of the simulated jets with velocities between 500~m/s and 1.8~km/s.
The cross-sections were calculated using the exact geometry of the observation derived with the help of SPICE kernels. 
We have determined that the lowest velocity needed in the case of the 2007 occultation is 1.2~km/s.
The maximum velocity in this case is not determined and can exceed this threshold.
We recognize that the actual criteria demand only the jets that participate in the creation of the narrowest features to have the determined vertical velocity while for the rest of the jets it is undetermined.
However, we use the same vertical velocity for all the modeled jets to restrict the number of free parameters for the DSMC model.

The geometry of the occultation is such that with this model setup, 39 jets form a linearly independent set -- a considerably smaller number compared to the solar occultation.
The exact selection of jets that participate in the linearly independent set can vary.
If there were no geometrical constrains, and 40 jets could be freely selected out of the set of 100, the number of possible combinations would be very large (of order $10^{27}$).
In our application the selection is determined by the geometry and thus the number of possible sets is reduced.
For example, jet number~14 can be swapped for jet number~39 and the resulting fit quality would not be affected by this swap ($\chi^2$-value not changing above sixth significant digit). 
If one desires to determine the participating jets with higher certainty, one needs to invoke the data from other sources, such as for example, thermal maps of the surface recorded by the CIRS instrument or data from in-situ instruments.

\begin{figure}[h!]
\begin{center}
\includegraphics[width=0.45\columnwidth]{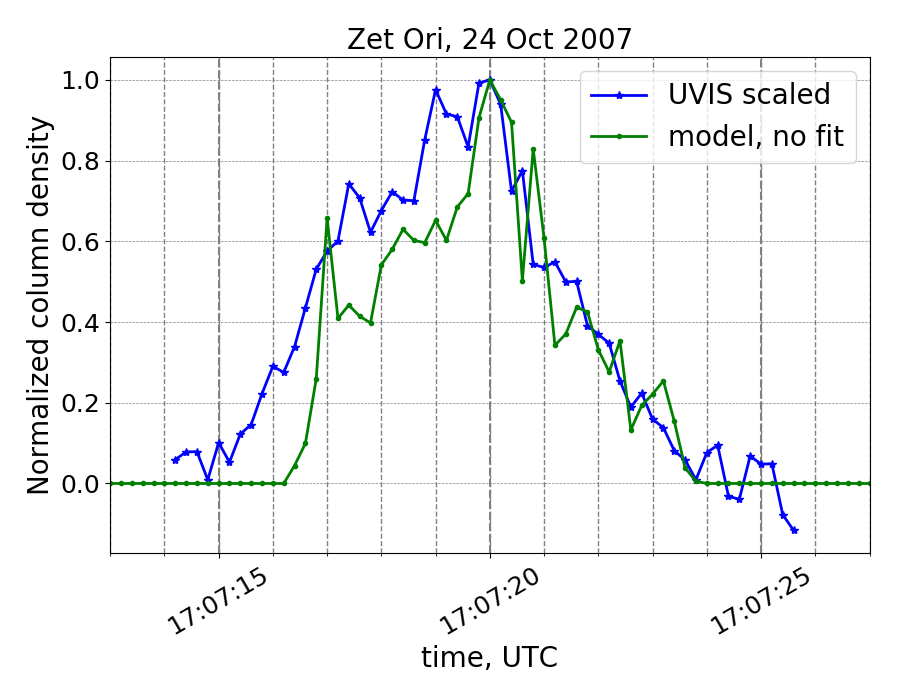}
\includegraphics[width=0.45\columnwidth]{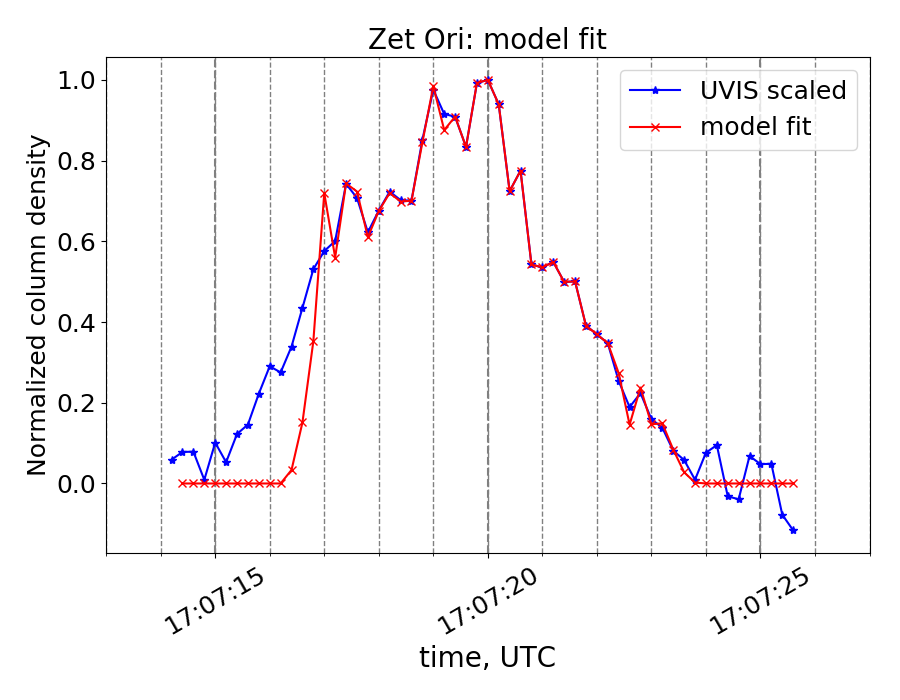}
\caption{\textbf{\label{fig:2007_fit}}
Comparison of modeled jets with UVIS occultation data from $\zeta$ Orionis occultation on 24 October 2007. Left panel: no fit model in which all 100 jets have coefficients $k_i=1$. Right panel: a possible fitted model with the 28 participating jets (their sources are shown in the right panel of Fig.~\ref{fig:2007_maps} ).}
\end{center}
\end{figure}

With 39 linearly independent jets and knowledge of how the UVIS LoS crosses them at each point in time, we can extract the synthetic column density and compare it to the observed one.
The green curve on the left panel of Fig. \ref{fig:2007_fit} shows the synthetic column density derived under the assumption that all jets are of the same strength.
The synthetic profile is shown in comparison with the observed column density derived from  HSP data shown in Fig. \ref{fig:2007_data}.
Both are normalized to unity for the sake of the fitting procedure and then plotted vs ephemeris time $et$. 
During the fitting, we vary only relative strengths of participating jets, while the absolute relation of modeled test particle to the real water vapor molecules comes from the comparison of total production rates (listed in Table \ref{tbl:occs_data}) after the fit is performed.
One can see that the shape of the no-fit synthetic profile generally coincides with the observational data and the largest observed enhancements in the attenuation signal naturally coincides with enhancements in the synthetic profile.
But the shape of the ramp of the data curve while getting into the occultation (ingress ramp) is more pronounced compared to the model and the need for a better fit over-all is noticeable.
Using the fitting procedure we then determined that out of 39 linearly-independent jets, 28 jets have non-zero contributions to the observed signal.
The resulting synthetic profile is plotted in the right panel of Fig. \ref{fig:2007_fit} .
The modeled jets now align better with the data in the mid-section of the occultation and in the egress while still failing in the ingress ramp.
This discrepancy between the UVIS profile and the modeled profile can be explained either by presence of non-orthogonal jets or by water vapor production from an extended area along the tiger stripes.
Both of these are deliberately excluded from our model to keep the number of fitting parameters to a minimum.

Fig.~\ref{fig:2007_maps} shows two possible sets of contributing jets.
These sets provide fits of comparable quality and serve as another representation of linear dependency arising during the transition from a three-dimensional water vapor distribution to the two-dimensional UVIS data.
As can be seen from the ground track (mapped in green squares) the closest approach of the UVIS LoS to the surface in this observation was over Baghdad.
It is logical to assume that the largest contribution to the attenuation signal is from jet sources at the Baghdad tiger stripe and thus the set plotted in the right panel is more suitable than the set from the left panel.
However, additional information, such as data from other instruments can be very useful to distinguish between the linearly-dependent jets. 
For example, thermal data could show that at the time of this UVIS occultation observation the Cairo tiger stripe was hotter than Baghdad. 
The observational mode for UVIS occultations is a rather specific one and the other instruments have difficulties observing Enceladus at the same time.
Thus the task of comparing of inter-instrumental observations turns into a separate project, similar to the work done by \cite{Teolis_2017}.
It is far beyond the scope of this paper. 

\subsection{$\epsilon$  and  $\zeta$ Orionis, 19 October 2011}

The UVIS occultation of 19 October 2011 was a very unique one because it was a double occultation: two stars from the Orion constellation were passing behind the jets and UVIS was able to follow the signal attenuation of both stars.
Two different spatial pixels of the UVIS FUV channel were used to follow both stars, $\epsilon$  and  $\zeta$ Orionis, while HSP port was pointed to $\epsilon$ Orionis.
For the purpose of fitting DSMC simulated jets, we used the data for $\epsilon$ Orionis because  although it is the dimmer of the two stars in the UV it cut through the plume at a lower altitude \citep{Hansen_2019}

\begin{figure}[h!]
\begin{center}
\includegraphics[width=0.98\columnwidth]{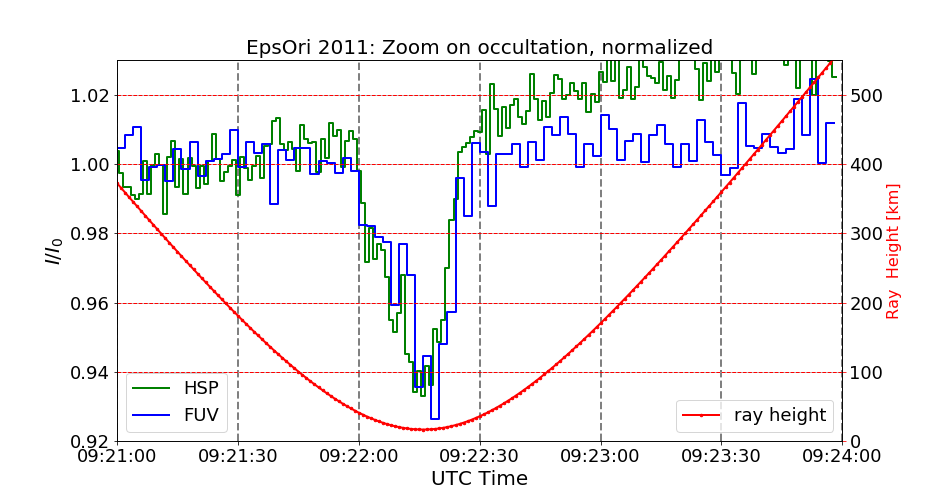}
\caption{\textbf{\label{fig:2011_data}}
HSP (green curve) and FUV (blue curve) data of the $\epsilon$ Orionis occultation from 19 October 2011. The red curve shows the ray height of the UVIS LoS with the closest approach altitude of 18~km.}
\end{center}
\end{figure}

\begin{figure}[h!]
\begin{center}
\includegraphics[width=0.45\columnwidth]{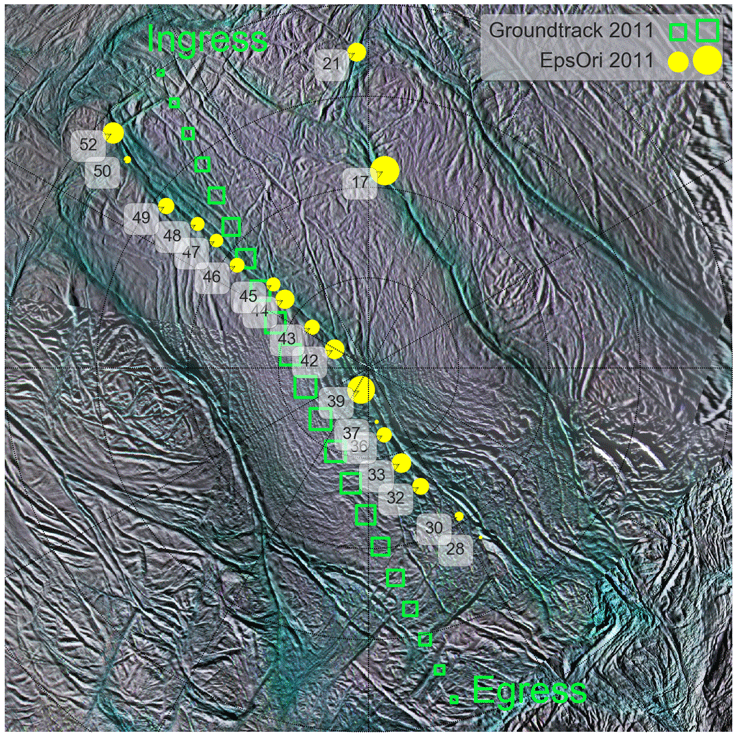}
\caption{\textbf{\label{fig:2011_map}} 
Locations of jet sources from the best fit to the data of $\epsilon$ Orionis occultation on 19 October 2011. }
\end{center}
\end{figure}

Fig. \ref{fig:2011_data} shows a plot of the occultation data for $\epsilon$ Orionis in the HSP and FUV channels together with the ray height plot. 
The maximum of absorption in this occultation coincides with the minimum of the ray height (18~km.)
The groundtrack for this observation can be seen in Fig. \ref{fig:2011_map} (green squares).
It was almost perfectly aligned with the orientation of the tiger stripes which means that the UVIS LoS at each point of this occultation was crossing jets of all tiger stripes. 
This causes the set of linearly independent jets to be rather small -- 30 model jets.
The fitting procedure using this set yielded 17 jets with non-zero coefficients.
The final fit and participating jet sources mapped on the map of southern polar regions are shown in Fig. \ref{fig:2011_map}.

\subsection{$\epsilon$ Orionis, 11 March 2016}

\begin{figure}[h!]
\begin{center}
\includegraphics[width=0.8\columnwidth]{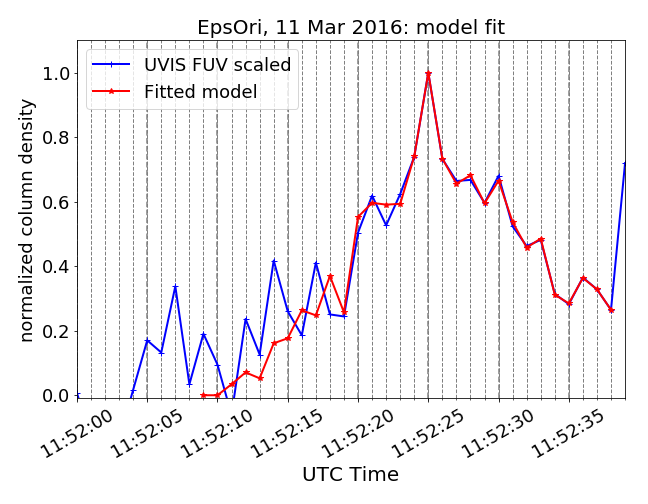}
\caption{\textbf{\label{fig:2016_fit}}
UVIS occultation observation of $\epsilon$ Orionis on March 11, 2016.}
\end{center}
\end{figure}

On 11 March 2016 UVIS observed $\epsilon$ Orionis passing behind the Enceladus plume for the second time. 
During this occultation, UVIS was able to observe only part of the plume because, unlike the occultation of 2011, the star passed behind the hard limb of Enceladus.
The occultation lasted at least 30~seconds and the FUV channel integration times were just one second. 
The ground track crossed mainly over Baghdad and Cairo slightly hitting Damascus on ingress while the LoS of UVIS hit the limb near Cairo. 
The highest absorption was detected when the LoS crossed over Baghdad at altitude of approximately 50~km.
Ground-track orientation and temporal resolution were generally favorable for distinguishing the jets arising from different tiger stripes with the restriction that just half of the polar region was observed.
The jet features in this occultation's profile (Fig. \ref{fig:2016_fit}) are rather wide and thus the vertical velocity of water vapor in the jets needed to fit them was smaller than in the the Solar occultation or in the later occultation of $\epsilon$~Canis Majoris in 2017. 
However, the absorption at times right before the LoS hit the limb dropped to just 2\% -- the same value as at 80~km during the ingress. 
This is an indication that the surface between the the tiger stripes emits little to no water vapor relative to the jets coming from the tiger stripes and another indication that the jets are highly collimated.

The Gaussian elimination for this occultation yielded 58 linearly independent model jets, which is a high number considering that only part of the polar region was observed.
We have randomly shuffled the jets to create several sets of linearly independent jets and run the fitting procedure on these different sets.
Out of 58, 28 jets came out to have non-zero coefficients.
Fig. \ref{fig:2016_maps} shows two sets of possible combinations of jets that provide a similar quality fits.
One can see that the majority of the jets are coming from the Baghdad tiger stripe in both examples while jets contributing to the fits from Cairo or Damascus are of relatively smaller strengths.
This was a general trend: the fits favor jets from Baghdad to those from Cairo or Damascus even at the closest approach near the end of occultation. 

\begin{figure}[h!]
\begin{center}
\includegraphics[width=0.95\columnwidth]{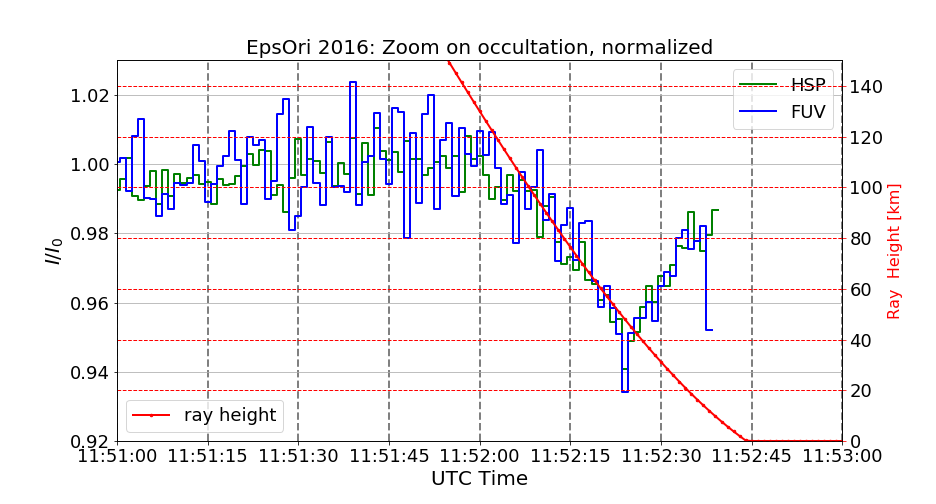}
\caption{\textbf{\label{fig:2016_data}}
UVIS occultation observation of $\epsilon$ Orionis on March 11, 2016.}
\end{center}
\end{figure}

\begin{figure}[h!]
\begin{center}
\includegraphics[width=0.45\columnwidth]{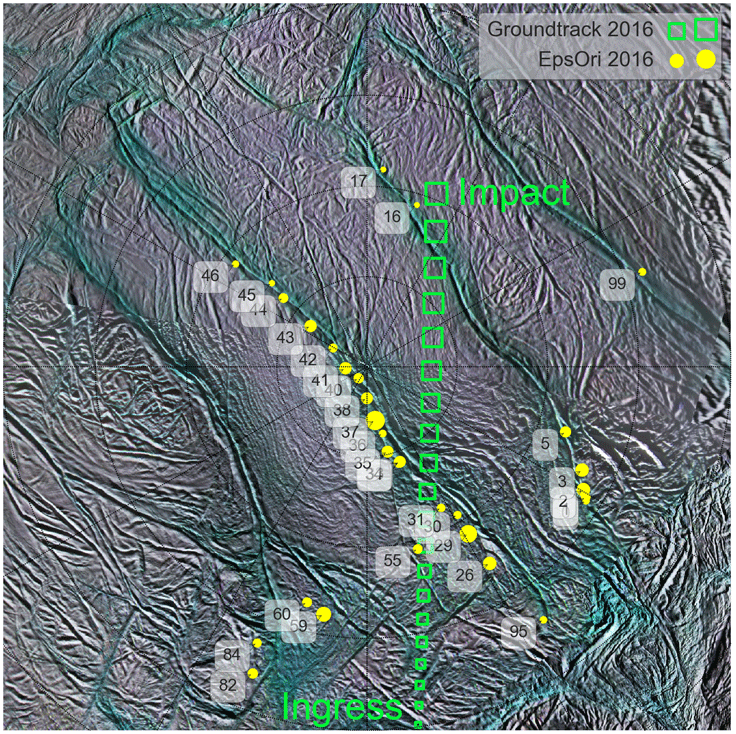}
\includegraphics[width=0.45\columnwidth]{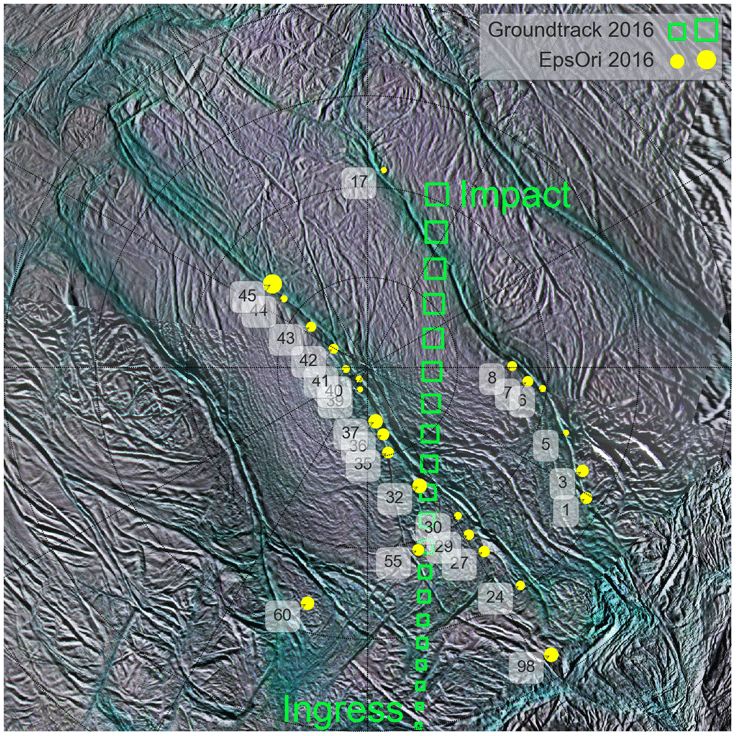}
\caption{\textbf{\label{fig:2016_maps}}
Locations of jet sources from two sets of jets. Both sets constitute a fit to the UVIS occultation observation of $\epsilon$ Orionis on March 11, 2016.
 (Fig. \ref{fig:2016_data}). Sizes of the circular markers indicate the relative strengths of each jet.}
\end{center}
\end{figure}

\subsection{$\epsilon$ Canis Majoris, 27 March 2017}

The last UVIS plume occultation observation of the Cassini mission was on 27 March 2017.
UVIS tracked $\epsilon$ Canis Majoris at a high altitude above the Enceladus surface in an unusual geometry: the groundtrack of this occultation was so far equatorial that it can not be mapped on the standard polar stereographic projection used in this paper.
The projection limits are at 75$^{\circ}$S and the ground track latitudes were around 60$^{\circ}$S.
This makes for a special oblique view to the plume: the LoS crossed all the tiger stripes simultaneously but the ray height became too great in the second half of the occultation, i.e. when the LoS was crossing over longitudes between 90$^{\circ}$W and 90$^{\circ}$E.
Similar to the occultation of 2016, we did not gain any new insights into that side of the plume from this occultation. 
The LoS orientation was similar to the occultation of 2011 with the difference that it crossed plume much higher and more obliquely.
The jet features indicated high velocity of the water vapor.
This means if we considered high-velocity well-collimated model jets, the linearly-independent set was large: 85 linearly independent jets.
The fitting procedure reduces this set to 32 jets with non-zero coefficients, which is still the second largest number after the Solar occultation.
The main reason for it is the short integration time of the FUV channel of just 1~second.

\begin{figure}[h!]
\begin{center}
\includegraphics[width=0.9\columnwidth]{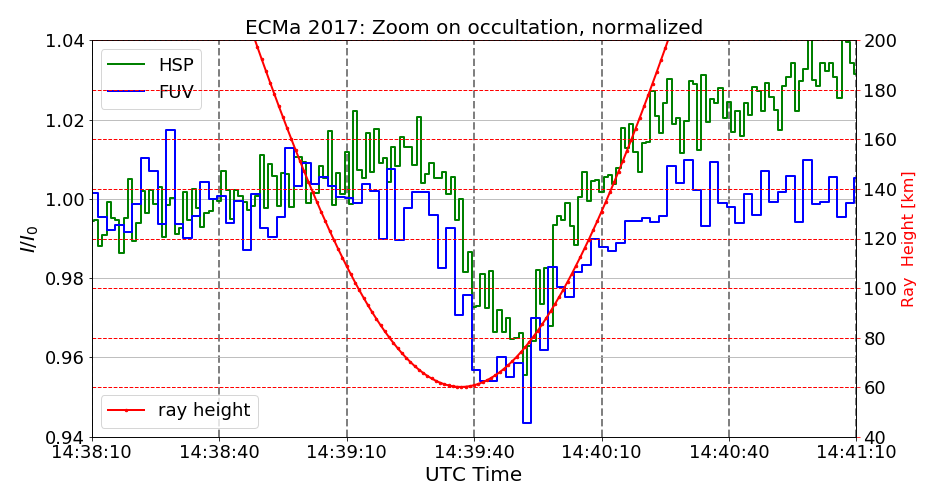}
\caption{\textbf{\label{fig:EOri2017_data}}
UVIS occultation observation of $\epsilon$ Canis Majoris on 27 March, 2017.}
\end{center}
\end{figure}

\begin{figure}[h!]
\begin{center}
\includegraphics[width=0.9\columnwidth]{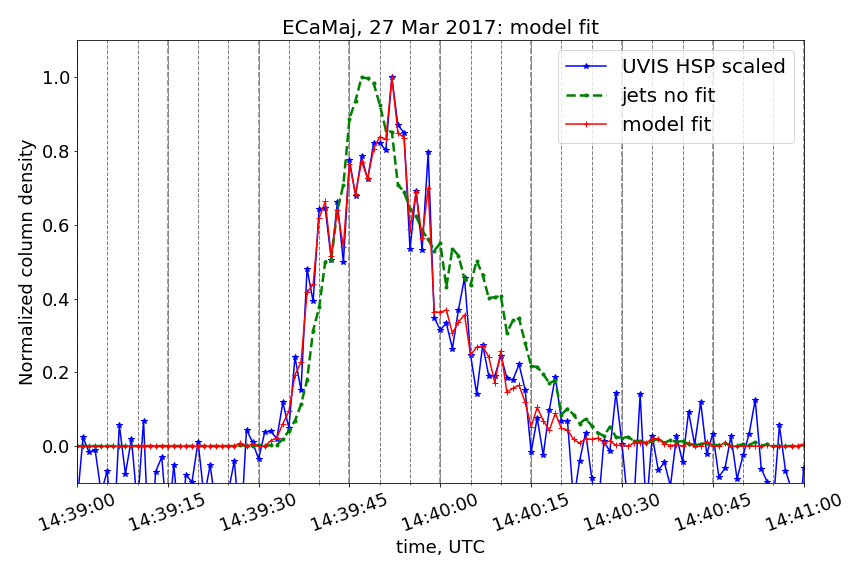}
\includegraphics[width=0.6\columnwidth]{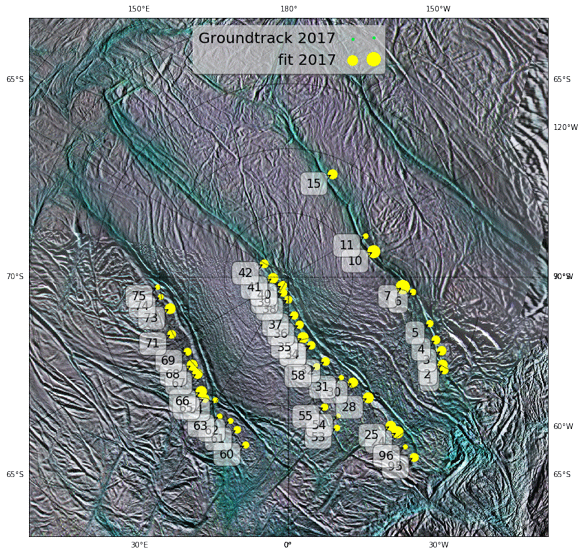}
\caption{\textbf{\label{fig:EOri2017_fit}}
Fit UVIS occultation observation of $\epsilon$ Canis Majoris on 27 March, 2017. Top panel shows best fit to normalized column density derived from UVIS HSP data. Bottom panel is the map of jet sources that contribute to the best fit. Relative size of dots reflects relative strengths of the jets. }
\end{center}
\end{figure}

\section{Conclusions}\label{sec:discussion}

In this work we have concentrated on modeling individual jets of pure water vapor.
The sensitivity of the data acquired by UVIS to the configuration of the jet sources, such as size and geometry of the opening of the vent, was beyond the scope of this work, however, it can be addressed with a similar procedure.
For example, alternative jet source configurations were proposed, such as "curtain sources" \citep{Spitale_2015}.
They can be modeled with minimal modification of the same DSMC model used in this work and the fit of such simulated jets to the UVIS data can be performed.

We have only considered jets erupting orthogonal to the surface.
This choice, while not necessarily realistic, was made deliberately to reduce the parameter space for the fitting procedure.
In 4 out of 5 occultations the procedure successfully found a good fit.

After examining all UVIS occultations  with the help of DSMC jet model fits it is apparent that, first of all, every one of them is special and yields different numbers of active jets and different combinations of the jets contributed to the best fits to UVIS data.
Table \ref{tbl:occs_fits} summarizes those UVIS occultations that were used to fit DSMC modeled jets.

UVIS observed different stars: $\epsilon$ Orionis in 2011 and 2016, $\zeta$ Orionis in 2007 and 2011, $\epsilon$ Canis Majoris in 2017, and the Sun in 2010. 
Due to different brightness in UV these stars produce different SNR.
With the strongest signal from the Sun, the solar occultation of 2010 had the highest SNR with the shortest integration times, and thanks to the beneficial geometry, allowed the model fit to identify 41 individual jets.
The integration times together with the height at which the LoS crossed the plume are the reason for the differences in the minimum vertical velocities required to fit the occultational profiles: it ranges from 1.2~km/s in the case of the 2007 occultation to 1.8~km/s for the Solar occultation of 2010.

The geometry of each occultation was different: UVIS made 4 horizontal cuts through the plume and 2 occultations in which the LoS hit the hard Enceladus limb. 
Out of 4 horizontal observations, in one the LoS was close to parallel to the tiger stripes (2010), in two the LoS cut across the tiger stripes (2007 and 2011), and in one the ground track was at non-polar latitudes, meaning the LoS crossedg the plume at high incidence and altitude (2017).
The geometry of 2011 occultation happened to be the least favorable to the fitting procedure because it resulted in the smallest linearly independent set of jets.
Jets from all tiger stripes were in the LoS of UVIS making it impossible to differentiate between them.
The geometry of the 2017 occultation, although similar to 2011, produced better results because the integration times were two times shorter.

The velocity of the Cassini spacecraft and consequently the spatial resolution of the occultations also varied: the occultations of 2010 and 2016 had the highest resolution across the ground track with 1~s integration time in EUV and FUV channels correspondingly.
2~ms integrations were used for HSP in 2007, 2016 and 2017.
In these occultations model fits identified a similar number of jets: 28, 29 and 32 respectfully.
Most of these jets participate in the best fit to all three of these occultations.

In 2016, the LoS hit the hard limb of Enceladus so that jets north of latitude approximately 87$^{\circ}$S and between longitudes 90$^{\circ}$W and 30$^{\circ}$E could have not contributed to the recorded signal.
In 2010, 2016, and 2017 the minimum of the H$_2$O absorption did not coincide with the closest distance of the LoS to the ground. 
In 2010 and 2016, this only can be explained by the difference between the jet strengths, while 2017 it may be because of the geometry.
The 2017 occultation had ra elatively high ray height above Enceladus' surface with a minimum of 60~km and the groundtrack hitting the surface at latitudes near 60$^{\circ}$S.
This resulted in an oblique view to the plume such that the jets in the longitude range from 90$^{\circ}$W to 45$^{\circ}$E were crossed at altitudes larger than 100~km and thus their contributions to the signal are small.

\begin{table}[h]
\begin{tabular}{ | l | l | l | l | l | }
\toprule
Occultation 						& Minimum   				& integration time		& V$_{min}$, 	& Nr. jets\\
										& 	ray height, km			& FUV(EUV)/HSP		&  km/s 			&  in fit\\
\midrule
$\zeta$ Ori, 2007 				& 12								& 5~s / 2~ms 			&  1.2				& 28 \\
Solar, 2010 						& 17								& 1~s / --					&  1.8				& 41	\\
$\epsilon$ Ori, 2011 		& 18								& 2~s / 8~ms 			&  1.3 				& 17	\\
$\epsilon$ Ori, 2016			& 0 (40)$^*$					& 1~s / 2~ms 			&  1.3				& 28	\\
$\epsilon$ CaMa, 2017 	& 60 							& 1~s / 2~ms 			&  1.8 				& 32	\\
\bottomrule
\end{tabular}
\caption{\textbf{\label{tbl:occs_fits}} Overview of UVIS occultation observations examined with fits of DSMC model jets. $^*$Second number corresponds to the altitude at which the Baghdad tiger stripe was crossed.}
\end{table}

We can conclude with certainty that 41 individual jets are required to fit the highest resolution UVIS dataset taken during the Solar occultation.
An alternative combination of a larger set of linearly-dependent jets can not be excluded.
A smaller number of jets are required to fit the stellar occultation data because of their resolution and geometry.

\begin{figure}[h!]
\begin{center}
\includegraphics[width=0.9\columnwidth]{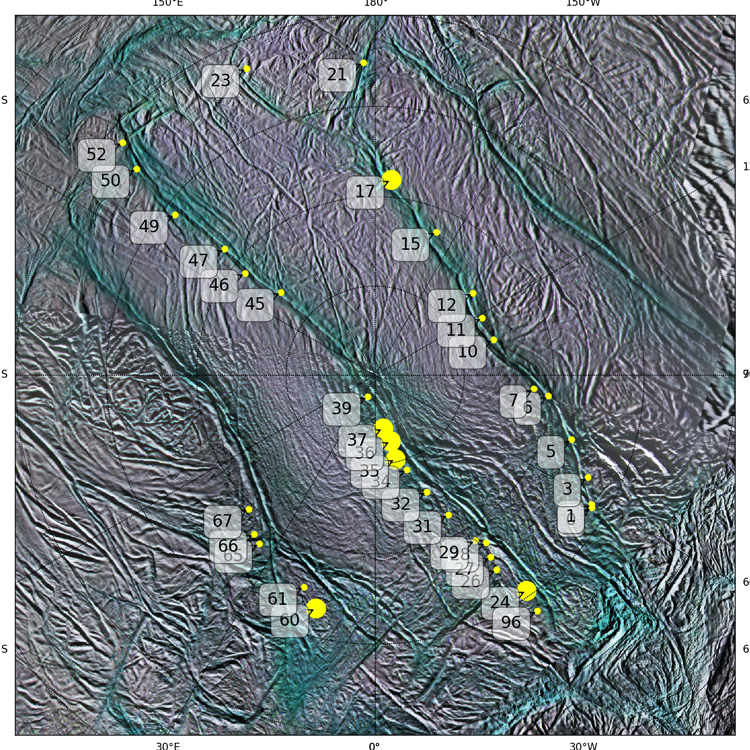}
\caption{\textbf{\label{fig:all_fits}}
Jets present in at least two of the occultation fits.}
\end{center}
\end{figure}

Fig. \ref{fig:all_fits} shows the locations of jet sources that are present in more than one best fit to UVIS occultation data.
This is a set of 37 jets that repeatedly appear in the best fits.
These jets overlap with the jets identified in \cite{Hansen_2011} despite at the time the authors used more coarse definitions for jets.
Most of jets from \cite{Hansen_2011} are represented by several jets plotted in Fig. \ref{fig:all_fits}:
in the nomenclature terms used in \cite{Hansen_2011}, jets numbers 65 to 67 correspond to Damascus II, Cairo V is represented by jets 6 and 7, Baghdad VI represented by the jet 45, jets 34 to 37 correspond to Baghdad VII, Cairo VIII is represented by jets 10 to 12.
It is highly probable that these jets were active constantly throughout the mission.

The combination of data from several instruments would yield new insights, for example help to eliminate some of the linear dependency between modeled jets.
Joint analysis of VIMS and UVIS data for the Solar occultation in 2010 and an investigation of temporal variability of jets using Cassini's UVIS and INMS observations can serve as an example of possible future studies \citep{Hedman_2018, Teolis_2017}.


\begin{landscape}
\section{Appendix A: Summary table of best fit coefficients to UVIS occultational data for 100 jets with source locations defined by Porco et al., 2014. }\label{tbl:all_fits_final_large_table}

Listed coefficients represent the best fit to each of UVIS occultation observations. 
For each occultation we list one possible combination of linearly independent jets.
Other combinations that include linearly-dependent substitutes to single jets or jet pairs, triplets, etc. are mathematically equivalent.  

Nan values indicate that corresponding jets were not a part of linear-independent set for the given occultation geometry. This effectively means that UVIS data does not include information on the strength of these jets. 
The coefficient equal to 0 indicates, that corresponding jets were part of linearly independent set and thus they must have contributed to UVIS LoS optical opacity. However, their contributions were determined by the fitting procedure to be negligible.

\begin{longtable}{llrrlrrrrr}
\toprule
   &   		&   		&    		&        			&       coefficient  &    coefficient     &     coefficient    &        coefficient  &        coefficient \\
{} &      ID &    lat &    lon &      Sulcus &  2007 &  2010 &  2011 &  2016 &  2017  \\
\midrule
\endhead
\midrule
\multicolumn{10}{r}{{Continued on next page}} \\
\midrule
\endfoot

\bottomrule
\endlastfoot
0  &  jet\_00 & -75.82 &  56.78 &       Cairo &        nan &       0.08 &       0.01 &        nan &        nan \\
1  &  jet\_01 & -75.93 &  57.43 &       Cairo &        nan &        nan &        nan &       0.40 &       0.27 \\
2  &  jet\_02 & -76.25 &  58.64 &       Cairo &        nan &        nan &       0.01 &       0.11 &       0.42 \\
3  &  jet\_03 & -76.81 &  62.59 &       Cairo &        nan &       0.10 &       0.01 &       0.30 &       0.35 \\
4  &  jet\_04 & -77.56 &  65.19 &       Cairo &       0.20 &        nan &        nan &       0.00 &       0.29 \\
5  &  jet\_05 & -78.44 &  70.08 &       Cairo &        nan &       0.22 &       0.05 &       0.29 &       0.16 \\
6  &  jet\_06 & -80.24 &  81.40 &       Cairo &       0.16 &        nan &       0.06 &       0.10 &       0.11 \\
7  &  jet\_07 & -81.08 &  83.31 &       Cairo &       0.13 &       0.18 &       0.15 &       0.00 &       1.00 \\
8  &  jet\_08 & -81.98 &  88.83 &       Cairo &       0.27 &       0.01 &       0.16 &        nan &        nan \\
9  &  jet\_09 & -83.08 &  98.01 &       Cairo &       0.15 &        nan &       0.22 &        nan &        nan \\
10 &  jet\_10 & -83.07 & 104.90 &       Cairo &       0.28 &       0.23 &        nan &        nan &       0.80 \\
11 &  jet\_11 & -83.20 & 116.30 &       Cairo &       0.32 &       0.05 &       0.28 &        nan &       0.06 \\
12 &  jet\_12 & -82.85 & 128.22 &       Cairo &       0.39 &       0.10 &       0.00 &        nan &        nan \\
13 &  jet\_13 & -82.55 & 135.22 &       Cairo &        nan &       0.03 &       0.32 &        nan &        nan \\
14 &  jet\_14 & -82.17 & 146.37 &       Cairo &       0.95 &       0.13 &        nan &        nan &        nan \\
15 &  jet\_15 & -81.29 & 155.01 &       Cairo &       0.25 &       0.07 &       0.00 &        nan &       0.40 \\
16 &  jet\_16 & -80.62 & 161.11 &       Cairo &       0.73 &       0.12 &       0.43 &        nan &        nan \\
17 &  jet\_17 & -79.04 & 173.53 &       Cairo &       0.82 &       0.09 &       1.00 &        nan &        nan \\
18 &  jet\_18 & -77.51 & 176.84 &       Cairo &        nan &        nan &       0.48 &        nan &        nan \\
19 &  jet\_19 & -76.72 & 178.43 &       Cairo &       1.00 &        nan &       0.00 &        nan &        nan \\
20 &  jet\_20 & -76.17 & 183.83 &       Cairo &        nan &       0.06 &       0.08 &        nan &        nan \\
21 &  jet\_21 & -72.59 & 180.39 &       Cairo &       0.82 &       0.00 &       0.28 &        nan &        nan \\
22 &  jet\_22 & -72.81 & 199.92 &       Cairo &       0.81 &       0.00 &       0.07 &        nan &        nan \\
23 &  jet\_23 & -71.51 & 201.00 &       Cairo &        nan &       0.09 &       0.00 &        nan &        nan \\
24 &  jet\_24 & -75.30 &  33.31 &     Baghdad &       0.49 &       0.01 &        nan &        nan &       0.68 \\
25 &  jet\_25 & -75.99 &  32.75 &     Baghdad &       0.18 &        nan &        nan &       0.70 &       0.47 \\
26 &  jet\_26 & -77.18 &  30.20 &     Baghdad &       0.38 &       0.34 &        nan &       0.30 &        nan \\
27 &  jet\_27 & -77.95 &  30.64 &     Baghdad &       0.06 &       0.05 &        nan &       0.00 &        nan \\
28 &  jet\_28 & -78.76 &  31.73 &     Baghdad &       0.12 &        nan &        nan &       0.00 &       0.53 \\
29 &  jet\_29 & -79.18 &  29.44 &     Baghdad &       0.18 &        nan &        nan &       1.00 &        nan \\
30 &  jet\_30 & -80.38 &  29.65 &     Baghdad &       0.16 &        nan &        nan &       0.00 &       0.44 \\
31 &  jet\_31 & -81.16 &  25.96 &     Baghdad &       0.18 &       0.23 &        nan &       0.10 &       0.04 \\
32 &  jet\_32 & -82.83 &  22.04 &     Baghdad &        nan &        nan &        nan &       0.51 &       0.28 \\
33 &  jet\_33 & -84.42 &  17.36 &     Baghdad &        nan &        nan &        nan &       0.39 &       0.07 \\
34 &  jet\_34 & -84.40 &  16.76 &     Baghdad &        nan &       0.07 &        nan &       0.00 &       0.26 \\
35 &  jet\_35 & -85.15 &  11.87 &     Baghdad &        nan &       0.17 &        nan &       0.00 &       0.54 \\
36 &  jet\_36 & -86.19 &  11.50 &     Baghdad &        nan &       0.07 &        nan &       0.32 &       0.24 \\
37 &  jet\_37 & -86.98 &   6.89 &     Baghdad &        nan &        nan &        nan &       0.44 &       0.24 \\
38 &  jet\_38 & -88.25 & 358.17 &     Baghdad &        nan &        nan &        nan &       0.30 &       0.22 \\
39 &  jet\_39 & -88.72 & 339.63 &     Baghdad &        nan &       0.13 &        nan &       0.00 &       0.24 \\
40 &  jet\_40 & -89.21 & 322.42 &     Baghdad &        nan &        nan &        nan &       0.00 &       0.32 \\
41 &  jet\_41 & -88.80 & 272.23 &     Baghdad &        nan &        nan &        nan &       0.38 &       0.46 \\
42 &  jet\_42 & -87.85 & 239.35 &     Baghdad &        nan &        nan &        nan &       0.10 &       0.28 \\
43 &  jet\_43 & -86.14 & 232.38 &     Baghdad &        nan &        nan &        nan &       0.27 &        nan \\
44 &  jet\_44 & -84.00 & 228.81 &     Baghdad &        nan &        nan &        nan &       0.20 &        nan \\
45 &  jet\_45 & -82.98 & 226.93 &     Baghdad &        nan &        nan &        nan &       0.02 &        nan \\
46 &  jet\_46 & -80.76 & 230.20 &     Baghdad &        nan &        nan &        nan &        nan &        nan \\
47 &  jet\_47 & -79.02 & 228.28 &     Baghdad &        nan &        nan &        nan &        nan &        nan \\
48 &  jet\_48 & -77.64 & 228.12 &     Baghdad &        nan &        nan &        nan &        nan &        nan \\
49 &  jet\_49 & -75.69 & 229.55 &     Baghdad &       0.38 &        nan &       0.08 &        nan &        nan \\
50 &  jet\_50 & -72.45 & 227.42 &     Baghdad &        nan &        nan &        nan &        nan &        nan \\
51 &  jet\_51 & -71.42 & 224.77 &     Baghdad &        nan &        nan &        nan &        nan &        nan \\
52 &  jet\_52 & -70.92 & 225.61 &     Baghdad &       0.32 &        nan &       0.25 &        nan &        nan \\
53 &  jet\_53 & -77.70 &  16.09 &     Baghdad &        nan &       0.08 &        nan &        nan &       0.09 \\
54 &  jet\_54 & -78.54 &  18.00 &     Baghdad &        nan &        nan &        nan &        nan &       0.01 \\
55 &  jet\_55 & -79.52 &  13.81 &     Baghdad &        nan &        nan &        nan &        nan &       0.17 \\
56 &  jet\_56 & -80.46 &  13.78 &     Baghdad &        nan &        nan &        nan &        nan &        nan \\
57 &  jet\_57 & -81.72 &  14.73 &     Baghdad &        nan &        nan &        nan &        nan &        nan \\
58 &  jet\_58 & -82.69 &  15.80 &     Baghdad &        nan &        nan &        nan &        nan &       0.14 \\
59 &  jet\_59 & -76.11 & 348.33 &    Damascus &        nan &       0.15 &        nan &        nan &        nan \\
60 &  jet\_60 & -76.56 & 343.99 &    Damascus &        nan &       0.09 &        nan &        nan &       0.14 \\
61 &  jet\_61 & -77.50 & 339.72 &    Damascus &        nan &       0.01 &        nan &        nan &       0.18 \\
62 &  jet\_62 & -77.96 & 336.33 &    Damascus &        nan &       0.09 &        nan &        nan &       0.06 \\
63 &  jet\_63 & -77.94 & 332.02 &    Damascus &        nan &        nan &        nan &        nan &       0.06 \\
64 &  jet\_64 & -78.88 & 327.56 &    Damascus &        nan &        nan &        nan &        nan &       0.06 \\
65 &  jet\_65 & -78.56 & 323.82 &    Damascus &        nan &       0.04 &        nan &        nan &       0.24 \\
66 &  jet\_66 & -78.82 & 321.01 &    Damascus &        nan &       0.11 &        nan &        nan &       0.58 \\
67 &  jet\_67 & -79.69 & 314.97 &    Damascus &        nan &       0.07 &        nan &        nan &       0.46 \\
68 &  jet\_68 & -79.86 & 310.73 &    Damascus &        nan &        nan &        nan &        nan &       0.53 \\
69 &  jet\_69 & -80.25 & 304.77 &    Damascus &        nan &        nan &        nan &        nan &       0.21 \\
70 &  jet\_70 & -80.31 & 302.85 &    Damascus &        nan &        nan &        nan &        nan &        nan \\
71 &  jet\_71 & -79.90 & 294.38 &    Damascus &        nan &        nan &        nan &        nan &       0.28 \\
72 &  jet\_72 & -80.32 & 291.57 &    Damascus &        nan &       0.14 &        nan &        nan &        nan \\
73 &  jet\_73 & -80.49 & 283.19 &    Damascus &        nan &        nan &        nan &        nan &       0.51 \\
74 &  jet\_74 & -79.97 & 277.06 &    Damascus &        nan &        nan &        nan &        nan &       0.05 \\
75 &  jet\_75 & -79.82 & 272.63 &    Damascus &        nan &        nan &        nan &        nan &       0.03 \\
76 &  jet\_76 & -77.95 & 263.80 &    Damascus &        nan &        nan &        nan &        nan &        nan \\
77 &  jet\_77 & -76.99 & 260.62 &    Damascus &        nan &        nan &        nan &        nan &        nan \\
78 &  jet\_78 & -76.39 & 255.44 &    Damascus &        nan &        nan &        nan &        nan &        nan \\
79 &  jet\_79 & -75.14 & 251.73 &    Damascus &       0.26 &        nan &        nan &        nan &        nan \\
80 &  jet\_80 & -73.14 & 249.07 &    Damascus &        nan &        nan &        nan &        nan &        nan \\
81 &  jet\_81 & -72.43 & 246.77 &    Damascus &        nan &        nan &        nan &        nan &        nan \\
82 &  jet\_82 & -71.97 & 337.85 &    Damascus &        nan &        nan &        nan &        nan &        nan \\
83 &  jet\_83 & -72.87 & 337.20 &    Damascus &        nan &       0.12 &        nan &        nan &        nan \\
84 &  jet\_84 & -73.60 & 336.59 &    Damascus &        nan &       0.02 &        nan &        nan &        nan \\
85 &  jet\_85 & -74.64 & 336.34 &    Damascus &        nan &       0.02 &        nan &        nan &        nan \\
86 &  jet\_86 & -75.55 & 333.09 &    Damascus &        nan &        nan &        nan &        nan &        nan \\
87 &  jet\_87 & -76.55 & 329.33 &    Damascus &        nan &        nan &        nan &        nan &        nan \\
88 &  jet\_88 & -75.07 & 116.16 &  Alexandria &        nan &       0.07 &        nan &        nan &        nan \\
89 &  jet\_89 & -74.56 & 153.20 &  Alexandria &        nan &       0.09 &        nan &        nan &        nan \\
90 &  jet\_90 & -73.17 & 156.72 &  Alexandria &        nan &       0.09 &        nan &        nan &        nan \\
91 &  jet\_91 & -70.56 & 164.46 &  Alexandria &        nan &       0.09 &        nan &        nan &        nan \\
92 &  jet\_92 & -69.58 & 165.51 &  Alexandria &        nan &       0.03 &        nan &        nan &        nan \\
93 &  jet\_93 & -74.10 & 184.81 &       Cairo &        nan &       0.18 &        nan &        nan &        nan \\
94 &  jet\_94 & -74.07 & 199.21 &       Cairo &        nan &        nan &        nan &        nan &        nan \\
95 &  jet\_95 & -72.99 &  33.13 &     Baghdad &        nan &        nan &        nan &        nan &       0.30 \\
96 &  jet\_96 & -74.04 &  32.78 &     Baghdad &        nan &        nan &        nan &        nan &       0.01 \\
97 &  jet\_97 & -75.95 & 131.55 &  Alexandria &        nan &        nan &        nan &       0.00 &        nan \\
98 &  jet\_98 & -71.24 &  30.85 &     Baghdad &       0.75 &        nan &        nan &        nan &        nan \\
99 &  jet\_99 & -73.94 & 107.27 &  Alexandria &        nan &        nan &        nan &       0.09 &        nan \\
\end{longtable}

\end{landscape}

\section{References:}
\bibliographystyle{elsarticle-harv}
\bibliography{anyas_references_arxiv.bib}


\end{document}